\documentclass{article}
\usepackage{spconf,amsmath,graphicx}

\usepackage[utf8]{inputenc} % allow utf-8 input
\usepackage[T1]{fontenc}    % use 8-bit T1 fonts
\usepackage{hyperref}       % hyperlinks
\usepackage{url}            % simple URL typesetting
\usepackage{booktabs}       % professional-quality tables
\usepackage{amsfonts}       % blackboard math symbols
\usepackage{nicefrac}       % compact symbols for 1/2, etc.
\usepackage{microtype}      % microtypography
\usepackage{xcolor}         % colors

\usepackage{algpseudocode}             % http://ctan.org/pkg/algorithmicx
\usepackage{algorithm}                 % http://ctan.org/pkg/algorithms
\graphicspath{ {./images/} } % configure image folder
\usepackage{import}         % seperate appendix files

\title{Mathematical Vocoder Algorithm : Modified Spectral Inversion for Efficient Neural Speech Synthesis }

\twoauthors
{Hyun Gon Ryu\sthanks{School of Mathematics and Computing,  Yonsei University }, Simon See }
{NVIDIA Coperation }
{Jeong-Hoon Kim}
{Yonsei University }

\begin{document}

\maketitle

\begin{abstract}
In this work, we propose a new mathematical vocoder algorithm(modified spectral inversion) that generates a waveform from acoustic features without phase estimation. The main benefit of using our proposed method is that it excludes the training stage of the neural vocoder from the end-to-end speech synthesis model. Our implementation can synthesize high fidelity speech at approximately 20 Mhz  on CPU and 59.6MHz on GPU. This is 909 and 2,702 times faster compared to real-time.  Since the proposed methodology is not a data-driven method, it is applicable to unseen voices and multiple languages without any additional work. The proposed method is expected to adapt for researching on neural network models capable of synthesizing speech at the studio recording level.
\end{abstract}

\section{Introduction}
Most neural end-to-end speech synthesis models(Shen et al., 2018, Ping et al., 2017, Li et al., 2019, Valle et al., 2019)   consist of a part that converts text into an acoustic feature and a part that converts the acoustic feature into a waveform. A vocoder converts acoustic features to a waveform. Recently, neural vocoders(Oord et al., 2016, Prenger et al.,2019 , Kumar et. al., 2019,  Kong et al.,2020, Kong et al.,2020) have been widely used to synthesize high fidelity voices. We study a new mathematical vocoder algorithm(new type of Spectral inversion) that generates a waveform from acoustic features without phase estimation. 

Since the proposed methodology is not a data-driven method, it is perfectly applicable to unseen voices and multiple languages. We evaluate our proposed algorithm for multi-speaker and multi-language datasets. Our implementation can synthesize speech at approximately 20 Mhz on an CPU  and 59.5Mhz on GPU without any hardware (H/W)-specific optimization. 

All reconstructed speech samples come from the ground truth acoustic features extracted from original speech. We will continue to work on voice synthesis from predicted acoustic features in follow-up studies. Our audio samples are available on the demo web-site.  \footnote{https://its10041004.github.io/MVA/}  In Section 4.3, we show the performance of spectral inversion comparing with neural vocoder Hifi-GAN(Kong et al., 2020).  Our implementation can synthesize high fidelity speech at approximately 20 Mhz on CPU and 59.6MHz on GPU. This is 909 and 2,702 times faster compared to real-time. We will update additional experimental results.

\section{Mathematical Vocoder Algorithm  }

\begin{table}[ht]
\centering
\small
\begin{tabular}{|l|c|l|l|l|}
\hline
Vocoder             & method       & trsf  & scale  & type        \\ \hline
Spec. Inversion  &  inversion     & FFT   & linear & complex          \\ \hline
Spec. Inversion  &  inversion     & FFT   & linear & mag\&phase     \\ \hline
GLA                 &  Phase Est.    & FFT   & linear & mag only         \\ \hline
WaveNet             & neural        & FFT   & Mel.   & mag only          \\ \hline
WaveGlow            & neural      & FFT   & Mel.   & mag only            \\ \hline
DiffWave            & neural         & FFT   & Mel.   & mag only         \\ \hline
Hifi-GAN            & neural      & FFT   & Mel.   & mag only            \\ \hline
Ours (Algo1 LF)     & inversion      & FFT   & linear & real             \\ \hline
Ours (Algo2 LD)     & inversion      & DCT   & linear & real             \\ \hline
Ours (Algo3 LP)     & inversion      & RFFT  & linear & packed \\ \hline
\end{tabular}
\caption{comparison of Vocoders : nueral vocoders use mel scaled one sided  magnitude only spectrogram from Fourier transfrom with win(1024) and hop(256).   }
\end{table}

In the standard STFT, The magnitude spectrogram is obtained by  modulus, or absolute values, of complex number $ z = x + i y $ which came from Fourier transform is defined as the nonnegative real number $ \sqrt{x^2 +y^2}$ and denoted by $|z|$ . It is same to measure the distance from the origin of the complex plane in the frequency domain using both the real and imaginary part. 

In Algorithm 1 , we uses only the  $x$-axis  value by removing the imaginary part of the complex plane. The real part $\textbf{Re} (z)$ of spectrogram obtained from standard STFT. Algorithm 2 and 3 replace DCT or RFFT instead of FFT transform. step2, zero clipping($\textbf{Re} (z)^+$)  is performed instead of absolute function. Zero clipping is expressed by the rectified linear(ReLU) activation function in neural network algorithms(Andrew et al.,2014). Zero clipping is a simple method($ \textbf{ReLU}(\textbf{Re} (z)) = \textbf{Re} (z)^+$). We also omit the zero clipping if the  text-to-acoustic feature model have capability to predict signed value. In Algorithm3 use both of real and imaginary part with packing algorithm with Hermitian (conjugate symmetric) properties of real input.  

\begin{algorithm}[h]
\small
%\SetAlgoLined
\begin{algorithmic}
\State   $  \textbf{input : x}  $ : waveform
\State  \quad \quad \quad $ S$ = STFT ($ \textbf{x}$ )     : Shift Time Fourier Transform   
\State  \quad \quad \quad $\textbf{R}$ = $\textbf{Real}(S) $  : real value
\State  \quad \quad \quad $\textbf{P}$ = $\textbf{R} ^{+} $   : zero clipping(optional)

\\\hrulefill
\State $ \textbf{input : P}$ : linear real zero-cliped Spectrogram    
\State  \quad \quad \quad $ \textbf{W}$ =iSTFT( $ \textbf{P}$ )   : inverse STFT  
\State $ \textbf{result : }  \textbf{W} $ reconstructed waveform
\end{algorithmic}
 \caption{ iSTFT for real zero clipped spectrogram }
\end{algorithm}

 \begin{algorithm}[h]
 \small
%\SetAlgoLined
\begin{algorithmic}
\State   $  \textbf{input : x}  $ : waveform
\State  \quad \quad \quad  $ \textbf{D}$ = STDCT ( $\textbf{x}$ )     : Shift Time DCT    
\State  \quad \quad \quad $\textbf{P}$ = $\textbf{D} ^{+} $   : zero clipping(optional)

\\\hrulefill
\State $ \textbf{input : P}$ : linear real zero-cliped Spectrogram    
\State  \quad \quad \quad $ \textbf{W}$ = iSTDCT ($ \textbf{P}$ )     : inverse STDCT
\State $ \textbf{result : }  \textbf{W} $ reconstructed waveform
\end{algorithmic}
 \caption{ iSTDCT for DCT spectrogram  }
\end{algorithm}

 \begin{algorithm}[h]
 \small
%\SetAlgoLined
\begin{algorithmic}
\State   $  \textbf{input : x}  $ : waveform
\State  \quad \quad \quad  $ \textbf{P}$ = STPRFT ( $\textbf{x}$ )     : Shift Time Packed Real Fourier Transform 
\State  \quad \quad \quad $\textbf{Z}$ = $\textbf{P} ^{+} $   : zero clipping(optional)
  
\\\hrulefill
\State $ \textbf{input : P} $ :  linear packed Spectrogram  
\State \quad \quad \quad  $ \textbf{W}$ = iSTPRFT ( $ \textbf{S}$ )     : inverse STPRFT
\State $ \textbf{result : }  \textbf{W} $ reconstructed waveform
\end{algorithmic}
 \caption{ iSTRFT for packed spectrogram }
\end{algorithm}

\subsection{Modified Spectrogram for Spectral Inversion }
\begin{figure}[h]

  \centering
  \fbox{
  %\rule[-.5cm]{0cm}{4cm} 
  \includegraphics[width=0.5\textwidth]{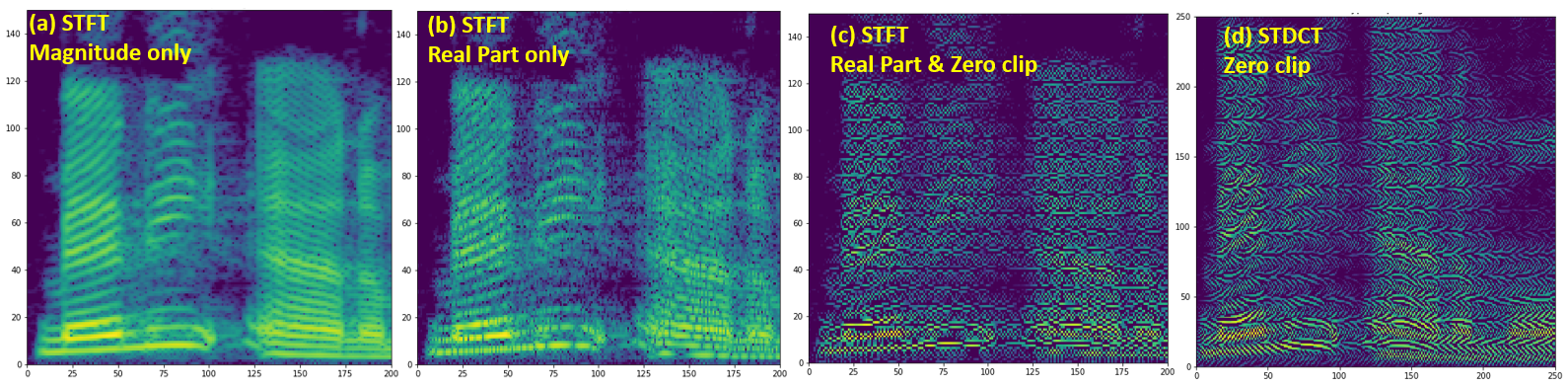} 
  %\rule[-.5cm]{4cm}{0cm}
  }
  \caption{ Comparison of the Spectrograms, magnitude only, real part only and zero clip }
  \label{fig:spec}  
\end{figure}

We could see a difference between the magnitude-only spectrogram and the real part-only spectrogram of STFT in Figure \ref{fig:spec}. Moreover, we could recognize a pattern after zero clipping.  In Algorithm 1, it's hard to synthesize high quality speech but it is our baseline to compare performance and quality with other existing methods. It is impossible to synthesize high-quality speech by using a mel scaled spectrogram with spectral inversion. For this reason, the modified spectrogram increases the amount of data to represent the acoustic feature and computation cost would be increase when we adapt proposed spectrogram in text-to-acoustic feature model.  

\section{Experiments}
\subsection{Dataset}

Since the proposed method is math-based, unlike other data-driven neural vocoders, there are no restrictions on data curation and data preparation procedures, such as separation of training data and test data. We selected LJSpeech (Ito, 2017) and LibriTTS(Zen et al., 2019) for English speakers and MLS(Pratap et al, 2020) for multi-language. The LJSpeech dataset has no restrictions on its use. The LibriTTS and  The MLS dataset are licensed under the CC BY 4.0.

\subsection{Computing Resources}

The equipment used in the experiment was NVIDIA DGX-1 with an Intel Xeon E5-2698 v4 @ 2.20GHz and NVIDIA Tesla V100 16GB. Since our proposed method is a mathematical algorithm(modified spectrial inversion), unlike neural vocoders, it does not require training infrastructures and can be applied directly to inference. Therefore, the equipment used in the experiment was also focused on verifying the performance of inference using a single core and Single GPU.

\subsection{Implementation}

All algorithms were implemented using the Python language. For easy verification and reproduction in different architectures, the public library was used as much as possible. Numpy, SciPy and Librosa library is commonly used in the CPU version. In the case of STFT/iSTFT of algorithm 1, there are various implementations such as torch built-in function, torchaudio, and torchlibrosa algorithm. In order to obtain the shift time transform, the convolution technique was applied after obtaining a DFT matrix or a fourier basis matrix in most implementations. OLA was applied to obtain the inverse transform. We exclude torch built-in STFT and iSTFT function for pair comparison because it utilize multicores in default. We also implement with scipy.fft/ifft for algorithm 1 and we replace  scipy.dct(type2,type3) instead of scipy.fft/ifft for Algorithm2 and scipy.fftpack.rfft/irfft for Algorithm3. The hann window from scipy was used for all cases. kaiser windows is also available. Most of cases, calculation of overlapping window was critical to reconstruct high quality audio to avoid artifacts.  The algorithm 3 with large hop size would be faster than algorithm 1 in computational perspective. However, our implementation of algorithm 3 is slower than algorhtm 1. When implementing STDCT and STRFT, frame processing was fully parallelized but iSTDCT and STRFT process was not fully parallelized to compute overlap. It is necessary to perform optimization later. 

We use GPU version of STFT/iSTFT for algorithm 1. For algorith 2 and 3, We prepared a naive CPU algorithm as baseline and performed GPU acceleration with cupy, JAX and pytorch. 

\section{Results}
We will update latest result for additional experiments for quality and performance. 

\subsection{Special Conditions}
We investigated when we could reconstruct a waveform without phase estimation from acoustic features directly. In the FFT case, it was confirmed that the artifact noises occurred frequently. In the DCT and RFFT case, these artifacts was not detected when the hop size was less than half of the win size. Since the speech synthesized by Algorithm 3 contains all the information necessary for restoration in a given spectrogram with packing, it could be restored with perfect sound quality under various conditions without any artifacts.

\subsubsection{Factors affecting sound quality}
The factor that most affects sound quality is hop size for Algorithm 2. It has been objectively verified that the sound quality of synthesized speech improves as the value of the hop size decreases. Since the speech synthesized by Algorithm 3 could be restored with perfect sound quality under various conditions without any artifacts. So we focus on Algorithm 1 and 2 for quality issues. 

\subsubsection{Denoising Effects}
In the case of the LJSpeech dataset, since the recording was performed in a non-studio environment, the noise could be checked in the silent section between active voices. It was confirmed that noise was removed from the restored sound source when clipping was performed based on a positive number less than 0.1 instead of zero clipping during the steps of generating spectrograms of Algorithm 2.  

\subsection{Quality of Synthesized Speech}
\subsubsection{MOS}
Mean opinion score (MOS) is a measure used in subjective  synthesized speech quality. MOS is expressed as a single rational number, typically in the range 1–5, where 1 is the lowest perceived quality and 5 is the highest perceived quality. MOS is calculated as the arithmetic mean over single ratings performed by human subjects for a given stimulus in a subjective quality evaluation test. MOS is widely used not only for the end-to-end TTS model but also for testing the sound quality of the synthesized voice of a vocoder.

To compare models, we report subjective listening test results rating speech naturalness and audio quality on a 5-point Mean Opinion Score (MOS) scale on Amazon Mechanical Turk. Our results can be seen in Table \ref{tab:mos_table}.

\begin{table}[]
\small
\centering
\caption{ Subjective Quality test of various models and compare R-MCD($ \downarrow $) and MOS1( $ \uparrow $ ) from Hifi-GAN paper(Kong et al. 2020) and MOS2(Ours) score. We will update additional experimental results.  }
\label{tab:mos_table}
\begin{tabular}{|l|r|r|r|}
\hline
vocoder            & R-MCD($ \downarrow $)  & MOS1($ \uparrow $)  & MOS2( $ \uparrow $ )  \\ \hline
GT                 & 5.93 & 4.45(0.06)    & 4.41 (0.10) \\ \hline
WaveNet            & 7.41 & 4.02(0.08)    & 3.88 (0.38) \\ \hline
WaveGlow           & 7.02 & 3.81(0.08)    & 4.07 (0.27) \\ \hline
MelGAN             & 7.49 & 3.79(0.09)    & 3.71 (0.31) \\ \hline
Hifi-GAN V1        & 6.94 & 4.36(0.07)    & 4.28 (0.10) \\ \hline
Hifi-GAN V2        & 7.41 & 4.23(0.07)    & 4.18 (0.08) \\ \hline
Hifi-GAN V3        & 7.31 & 4.05(0.08)    & 4.11 (0.19) \\ \hline
ours(LFZC 1024/64) & 6.99 & -             & 3.30 (0.20) \\ \hline
ours(LDZC 1024/64) & 6.28 & -             & 4.24 (0.10) \\ \hline
ours(LPS signed)  & 5.93 & -              & 4.25 (0.11) \\ \hline

\end{tabular}
\end{table}

\subsection{Performance of Spectral Inversion}
We used the default settings without performing any H/W-specific tuning such as boosting core clocks. The equipment used in the experiment was an Intel Xeon E5-2698 v4 @ 2.20GHZ and NVIDIA Tesla V100, and all measurements were made using a single core. For performance measurement, all runs were averaged after 100 runs with warm up. Data for all performance measurements were made using a 10-sec sound clip with a 22KHz sampling rate for convenience of measurement. To test the performance of the same conditions, we measure default windows size 1024, hop size 256 for reference. In Algorithm 2, a center pad was additionally performed to remove spike noise at both ends of the sound source. Since the required calculation amount varies according to the change of the window size and hop size, performance can be predicted based on the baseline window size 1024 and hop size 256.

\subsubsection{CPU performance}
In the environment (win length/hop size: 1024/256), our implementation of vocoder Algorithm 1 could synthesize speech at 4.5 Mhz with torchLibrosa implementation. It took 0.049 sec to generate a 10-sec audio clip. Our vocoder Algorithm 2 with DCT could synthesize speech at approximately 6.18 MHz, and it took 0.0357 sec. 

However,the reconstructed voice quality of Algorithm 1 and 2 depended on the win length and hop size. It was tested under various conditions. In Algorithm2, we could generate high fidelity speech with 1024/128. It could synthesize speech at 3.1 Mhz.  In Algorithm 3, we could reconstruct a waveform perfectly with any win size and hop size. We could configure special configurations, such as 1024/1022, 1024/768, 512/510 or 512/384, with which we used very coarse frames to fit in a number of phonemes or subwords in text embedding. Algorithm 3 is the same as Algorithm 2 in terms of performance with same windows and hop length. We could synthesize speech at  approximately 20 MHz with 1024/1022. It took 0.011 sec to generate a 10-sec audio clip. This is 909 times faster compared to real-time. 

\subsubsection{GPU acceleration}
By adopting this algorithm on GPU, it was possible to reduce the latency for the TTS service by connecting directly to the end-to-end TTS pipeline. The case of algorithm 1  were measured using torchlibrosa, the speech could be synthesized speed at 200 Mhz. For Algorithm 2 and 3, we implement naive DCT and RFFT is directly implemented, and OLA calculation is very slow because our implementation is in sequential. In the future, we plan to improve performance through parallel work. The algorithm 3 with large hop length can synthesize speech at 59.6 MHz with 1024/1022. It took 0.0037 sec to generate a 10-sec audio clip. This is 2,702 times faster compared to real-time. 

\begin{table}[]
\centering
\small
\caption{ comparison of inference speed of Vocoders on CPU and GPU. We are borrowing the perf. table from Hifi-GAN paper(Kong et al.,2020). Algo1 use torchlibrosa library.  Algorithm2 and Algorithm 3 use naive implementation and use cupy for GPU acceleration. We will update additional experimental results. }
\begin{tabular}{|l|r|r|r|r|}
\hline
            & CPU       & CPU      & GPU        & GPU       \\ \hline
 Vocoders   & KHz       & RTF(X)      & KHz        & RTF(X)       \\ \hline
 WaveGlow   & 4.7      & 0.2     & 501.0     & 22.8     \\ \hline
 MelGAN     & 145.5    & 6.6     & 14,238.0  & 645.7    \\ \hline
 Hifi-GAN V1& 31.7     & 1.4     & 3,701.0   & 167.9    \\ \hline
 Hifi-GAN V2& 215.0    & 9.7     & 15,863.0  & 764.8    \\ \hline
 Hifi-GAN V3& 296.4    & 13.4    & 26,169.0  & 1,186.8  \\ \hline
 Algo1(1024/256)& 4,515.7  & 204.8   & 200,454.6 & 9,090.9  \\ \hline
 Algo2(1024/256)& 6,176.5  & 280.1   & 1,721.3   & 78.1     \\ \hline
 Algo2(1024/128)& 3,132.1  & 142.1   & 864.4     & 39.2     \\ \hline
 Algo2(1024/ 64)& 1,404.5  & 63.7    & 429.4     & 19.5     \\ \hline
 Algo3(1024/1022)& 20,045.5 & 909.1   & 59,594.6  & 2,702.7  \\ \hline
 Algo3(512/510)  & 13,363.6   & 606.1   & 53,780.5  & 2,439.0  \\ \hline

\end{tabular}
\end{table}

\section{Benefits and Limitations }
\subsection{Benefit}
There are several benefits to adopting the proposed algorithm. 1) It eliminates the vocoder training part in the end-to-end TTS model. It is  also a universal vocoder for any unseen voice and multiple languages. 2) It is very fast in inference and saves computational cost and memory in inference. That means we could enhance the text-to-feature generation part while saving resources.  

\subsection{Limitation}
If we adopt our method in the end-to-end TTS model, we need to modify the current de facto standard log mel scaled mangitude only spectrogram-based end-to-end TTS pipeline with the proposed algorithm. When we apply the proposed algorithm  with  modified spectrogram directly to the end-to-end TTS pipeline, and it require more information to represent acoustic features. That means the computational cost(both of train and inference) increases in the text-to-acoustic feature model.

 \section{Discussion }
As shown, the proposed new vocoder algorithm(spectral inversion) without phase estimation.  We could reconstruct high fidelity speech from modified spectrogram.  In further research, we will evaluate our method with existing end-to-end neural TTS models. We believe that this will help in the deployment of high-quality audio synthesis, voice conversion, and music generation.  It is easily expanded to high-quality audio, such as CD or studio-quality audio, with a sampling rate of 44.1 KHz.  In addition, the modified spectrogram with Algoithm 3  can be used to create an acoustic token that can be completely restored like NLP. Through this, various follow-up studies are expected.

\section {References}

{
\small

Ahmed, Nasir; Natarajan, T.; Rao, K. R. (1974), "Discrete Cosine Transform". {\it IEEE Transactions on Computers}, C-23 (1): 90–93, doi:10.1109/T-C.1974.223784

Ahmed, Nasir (1991). "How I Came Up With the Discrete Cosine Transform".  {\it Digital Signal Processing. } (1): 4–5. doi:10.1016/1051-2004(91)90086-Z

Andrew L. Maas, Awni Y. Hannun, Andrew Y. Ng (2014). "Rectifier Nonlinearities Improve Neural Network Acoustic Models". {\it Proceedings of the 30 th International Conference on Machine Learning} 

D. J. Berndt and J. Clifford, "Using dynamic time warping to find patterns in time series." {\it in Proc. International Conference
on Knowledge Discovery and Data Mining}, 1994, pp. 359–370

Binkowski, M., Donahue, J., Dieleman, S., Clark, A., Elsen, E., Casagrande, N., Cobo, L. C., and Simonyan, K. (2019), "High fidelity speech synthesis with adversarial networks". arXiv preprint arXiv:1909.11646

JL Flanagan(1972), "Speech Analysis, Synthesis and Perception". {\it Springer- Verlag}, New York, 1972

Daniel Griffin and Jae Lim (1984). "Signal Estimation from Modified Short-Time Fourier Transform". {\it IEEE Transactions on Acoustics, Speech, and Signal Processing}, 32(2):236–243

De Cheveigne, A. and Kawahara, H. Yin, (2002), "A Fundamental Frequency Estimator for Speech and Music". {\it The Journal of the Acoustical Society of America}, 111(4):1917–1930

E. Jacobsen and R. Lyons(2003), "The sliding DFT". {\it Signal Processing Magazine vol. 20, issue 2}, pp. 74–80 

F. G. Meyer and R. R. Coifman (1997) "Applied and Computational Harmonic Analysis" 4:147

Gibiansky, A., Arik, S., Diamos, G., Miller, J., Peng, K., Ping, W., Raiman, J., and Zhou, Y. (2017), "Deep voice 2: Multispeaker neural text-to-speech". {\it In Advances in neural information processing systems}, pp. 2962–2970

Ito, K. et al. (2017), "The LJ speech dataset",  https://keithito.com/LJ-Speech-Dataset/ 

Jungil Kong, Jaehyeon Kim, Jaekyoung Bae (2020), HiFi-GAN: Generative Adversarial Networks for Efficient and High Fidelity Speech Synthesis,  arXiv preprint arXiv:2010.05646 

R. Kubichek, “Mel-cepstral distance measure for objective speech quality assessment”.  {\it in Proc. IEEE Pacific Rim Conf. on
Communications Computers and Signal Processing}, 1993.

John C. Steinberg (1937). "Positions of stimulation in the cochlea by pure tones". {\it Journal of the Acoustical Society of America.} 8 (3): 176–180. Bibcode:1937ASAJ....8..176S. doi:10.1121/1.1915891

Jont B. Allen (1977). "Short Time Spectral Analysis, Synthesis, and Modification by Discrete Fourier Transform". {\it IEEE Transactions on Acoustics, Speech, and Signal Processing.} ASSP-25 (3): 235–238. doi:10.1109/TASSP.1977.1162950

Lee, J., Choi, H.-S., Jeon, C.-B., Koo, J., and Lee, K.(2019), "Adversarially trained end-to-end korean singing voice synthesis system". arXiv preprint arXiv:1908.01919 

M. Morise, F. Yokomori, and K. Ozawa(2016), "WORLD: a vocoder-based high-quality speech synthesis system for real-time applications".  {\it IEICE transactions on information and systems}, vol. E99-D, no. 7, pp. 1877-1884  

M. Morise(2016), "D4C, a band-aperiodicity estimator for high-quality speech synthesis".  {\it Speech Communication}, vol. 84, pp. 57-65, Nov. 2016

M. Morise(2020), "Implementation of sequential real-time waveform generator for high-quality vocoder".  {\it in Proc. APSIPA ASC 2020}, pp. 821-825, Online, Dec. 7-10

Ping, W., Peng, K., Gibiansky, A., Arik, S. O., Kannan, A., Narang, S., Raiman, J., and Miller, J.(2017), "Deep voice 3: 2000-speaker neural text-to-speech". arXiv preprint arXiv:1710.07654

Prenger, R., Valle, R., and Catanzaro, B.(2019), "Waveglow: A flow-based generative network for speech synthesis". {\it In ICASSP 2019-2019 IEEE International Conference on Acoustics, Speech and Signal Processing (ICASSP)}, pp. 3617–3621. IEEE

Said, Amir; Pearlman, William A. (June 1996). "A new fast and efficient image codec based on set partitioning in hierarchical trees". {\it IEEE Transactions on Circuits and Systems for Video Technology.} 6 (3): 243–250. doi:10.1109/76.499834. ISSN 1051-8215

Saleem, N. Khattak, Muhammad Irfan Verdú, Elena (2019), "Spectral phase estimation based on deep neural networks for single channel speech enhancement". {\it  Journal of Communications Technology and Electronics} volume 64, pages1372–1382(2019)

Sejdić E.; Djurović I.; Jiang J. (2009). "Time-frequency feature representation using energy concentration: An overview of recent advances". {\it Digital Signal Processing. }19 (1): 153–183. doi:10.1016/j.dsp.2007.12.004

Shen, J., Pang, R., Weiss, R. J., Schuster, M., Jaitly, N., Yang, Z., Chen, Z., Zhang, Y., Wang, Y., SkerryRyan, R., et al.(2017), "Natural tts synthesis by conditioning wavenet on mel spectrogram predictions". {\it arXiv preprint} arXiv:1712.05884

Umesh, S. and Cohen, L. and Nelson, D. (1999), "Fitting the mel scale".  {\it Proc. ICASSP 1999}: 217–220, ISBN 978-0-7803-5041-0

Valle, R., Li, J., Prenger, R., and Catanzaro, B.(2019), "Mellotron: Multispeaker expressive voice synthesis by conditioning on rhythm, pitch and global style tokens". {\it arXiv preprint} arXiv:1910.11997

Vaidyanathan, P. P.(1993) , "Multirate Systems and Filter Banks", {\it Prentice Hall}, 1993.

Wang, Y., Skerry-Ryan, R., Stanton, D., Wu, Y., Weiss, R. J., Jaitly, N., Yang, Z., Xiao, Y., Chen, Z., Bengio, S., et al. (2017), "Tacotron: A fully end-to-end text-to-speech synthesis model". {\it arXiv preprint} arXiv:1703.10135 

Wang, Y., Stanton, D., Zhang, Y., Skerry-Ryan, R., Battenberg, E., Shor, J., Xiao, Y., Ren, F., Jia, Y., and Saurous, R. A.(2018), "Style tokens: Unsupervised style modeling, control and transfer in end-to-end speech synthesis". arXiv preprint arXiv:1803.09017
}

%\appendix
%\section{Appendix}

\end{document}